\documentclass[fleqn,prd,superscriptaddress,showpacs,nofootinbib,preprintnumbers]{revtex4}

\usepackage{amsmath}
\usepackage{amsfonts}
\usepackage{graphicx}
\usepackage{dcolumn}
\usepackage{fancyhdr}
\usepackage[mathscr]{euscript}
\def\hs{\qquad} 
\def\beq{\begin{eqnarray}} 
\def\eeq{\end{eqnarray}} 
\def\at{\left(} 
\def\aq{\left[} 
\def\ag{\left\{} 
\def\cp{\right.} 
\def\ct{\right)} 
\def\cq{\right]} 
\def\lap{\Delta\,} 

\def\segue{\qquad\Longrightarrow\qquad} 
\def\al{\alpha}

\def\ga{\gamma}
\def\de{\delta}

\def\ka{\kappa}

\def\si{\sigma}

\def\La{\Lambda}

\def\Om{\Omega}

\newcommand{\bea}{\begin{eqnarray}}
\newcommand{\eea}{\end{eqnarray}}
\newcommand{\beaa}{\begin{eqnarray*}}
\newcommand{\eeaa}{\end{eqnarray*}}

\def\nn{\nonumber}

\begin{document}

\title{\bf Topological electro-vacuum  solutions in extended gravity}

\author{
 Guido Cognola$\,^{(a)}$\footnote{cognola@science.unitn.it},
 Emilio Elizalde$\,^{(b)}$\footnote{elizalde@ieec.uab.es},
 Lorenzo Sebastiani$\,^{(a,c)}$\footnote{l.sebastiani@science.unitn.it },
 Sergio Zerbini$\,^{(a)}$\footnote{zerbini@science.unitn.it}
}
\affiliation{
$^{(a)}$ Dipartimento di Fisica, Universit\`a di Trento \\
and Istituto Nazionale di Fisica Nucleare \\
Gruppo Collegato di Trento, Italia\\
\medskip
$^{(b)}$ Consejo Superior de Investigaciones Cient\'{\i}ficas
(ICE/CSIC) \, and \\ Institut d'Estudis Espacials de Catalunya
(IEEC) \\
Campus UAB, Facultat Ci\`encies, Torre C5-Par-2a pl \\ 08193 Bellaterra
(Barcelona) Spain\\
$^{(c)}$Eurasian International Center for Theoretical Physics and Department of General \& Theoretical Physics, Eurasian National University, Astana 010008, Kazakhstan\\
}

\begin{abstract}
Spherically symmetric static topological black hole solutions
associated with some extended higher order gravitational models
in the presence of a Maxwell-field are derived by means of simple Lagrangian method, based on spherically symmetric reduction. Some new topological black hole solutions are presented and the validity of the First Law is investigated, and in these cases an expression for the energy is provided.
\end{abstract}
\pacs{04.50.Kd; 04.70.Dy; 97.60.Lf; 95.30.Sf}
\maketitle 

\section{Introduction}

Recent observational data seem to imply an accelerating expansion of the
visible universe leading to the so called Dark Energy issue.
There are several possible descriptions of  this acceleration.
Among them, the simplest one consists in introducing a small positive
cosmological constant in the, up to now, unchallenged framework of
General Relativity (GR), the so called $\Lambda$CDM model.
A generalization of this simple modification of GR consists in considering
modified gravitational theories,
where some combination of curvature invariants (the Riemann tensor, the Weyl tensor, the Ricci tensor and so on) replaces, or is added, to the classical Hilbert-Einstein action build on the Ricci scalar term $R$ only.
The simplest class of modified theories is $F(R)$-modified gravity, in which the action is described by a function $F(R)$ of the Ricci scalar $R$
(for a review see Refs.~\cite{review6,review7}).
Typically, these modified models admit de Sitter space as
a solution, its stability having been investigated by several authors
(see, e.g., \cite{guido,guido2,Far,Monica}). Furthermore, viable $F(R)$ models,
namely those which are able to pass the local gravitational GR tests and also
to give rise both to inflation and to the dark energy epoch in a unified way,
have been extensively discussed \cite{Saw,seba10,Od1,Od2}.
Another interesting class of modified gravitational models, with roots in the principles themselves of gravitational theories, is given by Weyl's conformal gravity, where the quadratic Weyl scalar appears. We would like to stress that indeed the detailed study of Weyl gravity (as e.g. its role in Lovelock's theories~\cite{lovelock}, vacuum energy, dark matter, etc.) may definitely improve our understanding of these cosmological problems.

One should note that the current acceleration of the universe is actually a low-curvature phenomenon, while it is usually recognized that black holes are associated with high-curvature effects. However, as stressed in Ref. \cite{nunes} (see also the references therein), the investigation of static spherically symmetric exact solutions in modified gravity models may have real interest also from the observational point of view with regard to strong gravitational lensing, where they could lead to some observable effects beyond those of GR, which could be used to prove or disprove these theories.

In the present paper we focus our interest on the impact of these theories on the physics of the corresponding black holes (BH). Static, spherically symmetric (SSS) solutions describing static black holes in general gravitational theories have been investigated in several papers. Typically, modified models
admit de Sitter space as a solution, but the issue to find exact
SSS metrics different from the Schwarzschild-dS one appears to be a formidable task since,
already for very reasonable models, the ensuing equations of motion are much more complicated than the corresponding ones for GR. For that reason the number of exact non-trivial
static black hole solutions known so far in modified theories of gravity is extremely small (see Refs.~\cite{CB,Deser,Multamaki,CapozzielloStabileTroisi,Bezerra,Saffari,SSSsolutions}).
Furthermore, these black hole solutions are not expected to share the same
laws of their Einsteinian counterparts. In parallel with ordinary BH solutions,
among the physical quantities one
should assign to these modified gravity black holes are a mass, horizon entropy, temperature, and so on. In particular,
the issue of associating an energy (mass) to these black hole solutions is quite problematic
(see Ref.~\cite{Visser:1993nu}; in M. Visser's terminology, most of the BH solutions in modified gravity theories are in fact ``dirty black holes''), and the modern terminology usually refers to them as Lifshitz BH solutions.

Several attempts addressed to find a satisfactory answer to the mass problem
have appeared (see, for instance, Refs.~\cite{D,Visser,Cai}, and references therein), but the definition of a sensible mass parameter is still a much debated question.
In Ref.~\cite{SSSEnergy}, for a particular class of BH solutions,  an identification for the BH energy has been proposed,
as a quantity proportional to the unique constant of integration which appears
in the explicit solutions of $F(R)$-modified gravity models in vacuum.
This identification was achieved by making use of a derivation of the First Law
of black hole thermodynamics from the equations of motion of $F(R)$-gravity,
by independently evaluating the entropy, via the Wald method~\cite{Wald}, and the
Hawking temperature~\cite{HT}, via quantum mechanical methods in curved space-times.
Some explicit examples along these lines have already been discussed in Ref.~\cite{SSSEnergy}.

Here we go further into this analysis by considering a particular class of (topological)
SSS solutions in presence of Maxwell field \cite{habib}, and this inclusion
is investigated in $F(R)$-gravity and in Weyl conformal gravity models.
With regard to this issue, we note that the Maxwell action and equations read, respectively,
\begin{equation}
I_{EM}=-\frac{1}{4}\int_{\mathcal{M}} d^4 x\sqrt{-g}\,F^{ij}F_{ij}\,
\,,\hs\hs \nabla_k F^{kj}=0\,,
\label{action0em}\end{equation}
where $j,j,k,...$ run from $0$ to $3$, $g$ is the determinant of the metric tensor $g_{ij}$,
$\mathcal{M}$ the space-time manifold, ${\nabla}_{i}$ is the covariant derivative
operator associated with $g_{i j}$ and
$F_{ij}$ the electromagnetic field strength.

We will look for static, (pseudo)-spherically symmetric solutions (SSS)
with various horizon topologies, thus the  generic metric reads
\begin{equation}
ds^2=-e^{2\alpha(r)}B(r)dt^2+\frac{d r^2}{B(r)}+r^2\,d\si^2_k\,,\hs\hs
d\si^2_k=\frac{d\rho^2}{1-k\rho^2}+\rho^2 d\phi^2\,,
\label{metric0}\end{equation}
where $\alpha(r)$ and $B(r)$ are functions of $r$ only. This form of the metric is chosen in order to deal with static
spherically symmetric BH solutions. We recall that one has a BH solution if there exists a simple positive zero
of $B(r_H)=0$, with $^{2\alpha(r_H)}$ being finite and non vanishing. Dirty or Lifshitz BH solutions correspond to
$^{2\alpha(r_H)} \neq 1$.  Where not necessary, the argument of all these functions  will be dropped. Furthermore, one has a topological BH if the ``horizon'' manifold with metric $d\si^2_k$ is a sphere $S_2$, a torus $T_2$,
 or a compact hyperbolic manifold $Y_2$, according to whether $k=1,0,-1$, respectively.

Due to the spherical symmetry, it is easy to show that the non vanishing
components of the electromagnetic field are
\beq
F_{01}=F_{tr}=\frac{e^{\alpha(r)}Q}{r^2}\,,\hs\hs
   F^{01}=-\frac{e^{-\alpha(r)}Q}{r^2}\,,
\eeq
$Q$ being the electric charge.
As a result, within the SSS Ansatz, the Maxwell action simplifies to
\begin{equation}
I_{EM}=\frac{V_k}{2}\int dt \int dr \frac{e^{\alpha(r)}Q^2}{r^2}\,,
\label{action1em}
\end{equation}
where $V_k$ is the volume of the ``horizon'' manifold, namely $V_1=4\pi$ (the sphere),
$V_0=|\Im\,\tau|$, with $\tau$ the Teichm\"{u}ller
parameter for the torus, and finally
$V_{-1}=4\pi g$, $g>2$, for the compact hyperbolic manifold with genus $g$ (see, for example \cite{Vanzo}).


The paper is organized as follows.
In Sect.~II we obtain SSS solution in topological Maxwell-$F(R)$-gravity.
In Sect.~III we extend to the electro-vacuum case the Weyl gravity topological SSS solutions of Refs.~\cite{r,m,klem} to observe that, also in this case, the solutions admit three free integration constants.
In Sect.~IV we discuss the existence conditions of BH event horizons.
Sect.~V is devoted to the evaluation of Wald's entropy in the well-know case of $F(R)$-gravity and in the non trivial case of Weyl gravity. In this Section we also present a simple proof that the conformal invariance of Wald's entropy for conformally invariant gravity models.
In Sect.~VI,  the topological version of the First Law of BH thermodynamics for $F(R)$-gravity is presented, and
a discussion regarding Weyl's conformal gravity is given. Conclusions are provided in Sect.~VII. Finally, in the Appendix, a simple proof of the conformal invariance of the Killing static spherically symmetric BH temperature is presented.
Throughout the paper
we use units of $k_{\mathrm{B}} = c = \hbar = 1$ and denote the
gravitational constant $G_N$ by $\kappa^2\equiv 8 \pi G_{N}$, so that
$G_{N}^{-1/2} =M_{\mathrm{Pl}}$, being
$M_{\mathrm{Pl}} =1.2 \times 10^{19}$GeV the Planck mass. Furthermore, we will also put $2\kappa^2=1$, for simplicity.

\section{Topological SSS solutions in Maxwell-F(R)gravity}
\label{S:SSSsol}

The  Maxwell-F(R)gravity  model is described by the action $I=I_R+I_{EM}$,
where the action for modified $F(R)$-theories reads
\begin{equation}
I_R=\frac{1}{2\kappa^2}\int_{\mathcal{M}} d^4 x\sqrt{-g}\,F(R)
\segue I=\int_{\mathcal{M}} d^4 x\sqrt{-g}\,\aq F(R)-\frac14\,F^{ij}F_{ij}\cq
\label{action0}\end{equation}
(as advanced, we are useing units in which $2\ka^2=1$).
$F(R)$ is an arbitrary smooth function of the Ricci scalar $R$ which, using the metric
in (\ref{metric0}), assumes the form
\begin{eqnarray}
R&=&-B''-\frac{B'}{r}\,(4+3r\al')-\frac{2B}{r^2}\,[(1+r\al')^2+r^2\al'']
          +\frac{2k}{{r}^{2}}\,,
\label{R}\end{eqnarray}
where here and in the following the prime $'$
represents derivation with respect to the $r$ variable.

The equations of motion for the action in (\ref{action0}) read
\beq
(g_{ij}\lap-\nabla_i\nabla_j+R_{ij})\,\frac{dF(R)}{dR}-\frac12\,g_{ij}\,F(R)=
-\frac18\,g_{ij}\,F^{rs}F_{rs}+\frac12\,F_{ir}F_{js}\,g^{rs}\,.
\label{fieldEq}\eeq
Of course, only two of the latter field equations are independent since the metric
depends on two arbitrary functions $\al(r)$ and $B(r)$, only.
One has
\begin{eqnarray}
F(R)=Rf-\frac{2f}{r^2}\,(k-B-rB')+\frac{f'}{r}\,(4B+rB')+2Bf''+\frac{Q^2}{2r^4}\,,
\label{one}
\end{eqnarray}
\begin{eqnarray}
\frac{\al'}{r}\,(2f+rf')-f''=0\,,
\label{two}\end{eqnarray}
where, for the sake of simplicity, we have set
\beq
f=f(r)=\frac{dF(R)}{dR}\,,\hs\hs f'=\frac{df}{dr}=\frac{dR}{dr}\,\,\frac{d^2F(R)}{dR^2}\,,\hs\hs
\mbox{and so on}\,.
\nn\eeq
In principle, given the function $F(R)$, the above equations permit to derive
both $\al(r)$ and $B(r)$, that is, the explicit form of the metric.
Alternatively, fixing the form of $\al(r)$ one derives $B(r)$ and,
as a consequence, the values of $F$ and $R$ as functions  of $r$
evaluated on the solution.
From such expressions one can ``try to reconstruct'' the form of $F$ as a function of $R$.
In general, the Lagrangian $F(R)$ one eventually finds is not unique, since one has to
infer its form starting from the value it assumes on the solutions.
For example, all Lagrangians of the form $F(R)=Rf(R)$ with
$\lim_{R\to0} Rf(R)=0$ have the Schwarzschild solution.
 
A comment is here in order. It has to be stressed that the general solutions of (\ref{one}) and (\ref{two}) are in fact physically acceptable only for some values of the parameters. In the following, even if not specified, it is always understood that all free parameters have been restricted to
values which give rise to physical solutions.

Just to see how the  reconstruction procedure works,
we first take the derivative of equation (\ref{one}) with respect to $r$, thus obtaining
\beq
&&\frac{\al'}{r^2}\aq rB'(4f-rf')+ B(-4f+2r^2f'')\cq
               -2(\al')^2Bf'+\frac{4\al''Bf}{r}
\nn\\ &&\hs\hs  +
\frac{f}{r^3}\,[4k-4B+2 B''r^2]
+\frac{f'}{r^2}\,[2rB-4 B)
+\frac{f''}{r}\,[4B+rB']-\frac{2Q^2}{r^5}=0\,,
\label{tre}\eeq
where, in order to simplify the result,  we have also used Eqs.~(\ref{R}) and (\ref{two}).
The usefulness of the latter equation is due to the fact that it only depends on $f$
which, in principle, can be derived from (\ref{two}), once $\al(r)$ is fixed.

Here we shall consider two important cases, that is, $\al(r)=$constant and
$\al(r)\sim\log(r^z)$ (of Lifshitz type, but in this case we will work with $Q=0$). In the first case the metric reads
\begin{equation}
ds^2= -B(r)dt^2+\frac{d r^2}{B(r)}+r^2\,d\si^2_k\,,
\label{metric00}
\end{equation}
and equations (\ref{R})-(\ref{tre}) notably simplify.
In particular, from (\ref{two}) it follows that
\begin{equation}
f''(r)=0\segue f=ar+b\,,\hs\hs a,b=constant\,.
\label{due2}\end{equation}
Putting this result into (\ref{tre}) and integrating, one gets
\begin{equation}
\begin{array}{lll}
1)&a=0,b\neq0\,,&\hs
 B(r)=k+\frac{c_1}{r}+c_2r^2+\frac{Q^2}{4br^2}\,,\\
2)&a\neq0,b=0\,,&\hs
 B(r)=\frac{k}2+\frac{c_1}{r^2}+c_2r^2+\frac{Q^2}{5ar^3}\\
3)&a,b\neq0\,,&\hs
 B(r)=\frac{k}{2}+\frac{bk}{3ar}+c_2\,r^2+\frac{Q^2}{4br^2}
       +c_0\,\at\frac{b^2}{2a^2}-\frac{br}{a}-\frac{b^3}{3a^3r}
+r^2\,\,\log\frac{ar+b}{ar}\ct\,,
\end{array}
\label{Bab}\end{equation}
$c_0,c_1,c_2$ being arbitrary integration constants.
The last expression is the more general form which the $B(r)$ function can assume
if the metric (\ref{metric00}) is a solution of a generic $F(R)$ theory with
electromagnetic minimal coupling. We see that it depends, at most, on
four independent parameters ($a,b,c_0,c_2$).

The corresponding scalar curvatures are given by
\begin{equation}
\begin{array}{lll}
1)&a=0,b\neq0\,,&\hs R=-12c_2\,,\\
2)&a\neq0,b=0\,,&\hs R=-12c_2+\frac{k}{r^2}-\frac{2Q^2}{5ar^5}\,,\\
3)&a,b\neq0\,,&\hs
R=-12c_2+\frac{k}{r^2}+c_0\,\aq-12\,\log\frac{ar+b}{ar}
        +\frac{b(12a^3r^3+4ab^2r+18a^2br^2-b^3)}{a^2r^2(b+ar)^2}\cq\,.
\end{array}
\label{Rab}\end{equation}
Now, in some cases we are able to ``infer'' the function $F(R)$ which generates
the solutions in (\ref{Bab}).
In the first case ($a=0,b\neq0$),
using (\ref{Bab}), (\ref{Rab}) and (\ref{one}) one may  recover  GR 
\beq
a=0,b\neq0\,,\hs\hs
\ag\begin{array}{l}
   B(r)=k-\frac{c_1}{r}-\frac{\La r^2}3+\frac{Q^2}{4 r^2}\,,\\
   R=4\La\,,\\
   F(R)=(R-2\La)=\,,
\end{array}\cp
\label{deS1}\eeq
where we have set $c_2\to-\La/3$ and $b\to1$. This is the topological Reisnner-Nordstrom solution with non
vanishing cosmological constant. We stress that the latter equation represents the value of $F(R)$ evaluated on the
solution. Of course, there are infinitely many functions $F(R)$ which reduce (on shell) to
$F(R)=2\Lambda$ for $R=4\La$ and, in general, they do not give the correct solution.


In the second case here considered, that is $a\neq0,b=0$, we have
\beq
a\neq0,b=0\,,\hs\hs
\ag\begin{array}{l}
 B(r)=\frac{k}2+\frac{c_1}{r^2}+c_2r^2+\frac{Q^2}{5ar^3}\\
R=-12c_2+\frac{k}{r^2}-\frac{2Q^2}{5ar^5}\,,\\
F(R)=\frac{2ak}r-\frac{Q^2}{2r^4}\,,
\end{array}\cp
\label{mod2}\eeq
where $a,c_1,c_2$ are arbitrary constants.
In the special cases $Q=0$ or $k=0$, namely in the absence on electromagnetism or in the flat case,
we are able to get $r$ as a function of $R$.

If $Q=0$, we get
\beq
r=\sqrt{\frac{k}{R+12c_2}}\segue F(R)=2a\,\sqrt{k(R+12c_2)}\,.
\label{deS2}\eeq
The latter function satisfies the condition (\ref{due2}) and so we can try to use it in order to
build up the action. One can directly verify, using (\ref{fieldEq}), that the action
\beq
I=\int_{\cal M}\,\sqrt{-g}\,d^4x\,\aq 2a\sqrt{k(R+12c_2)}\cq\,,\hs\hs k\neq 0\,,\label{es}
\eeq
has a vacuum solution of the form (\ref{metric00}), with
\beq
B(r)=\frac{k}2+\frac{c_1}{r^2}+c_2r^2\,.
\label{esB}\eeq
In a similar way, if $k=0$ we have, with $a<0$,
\beq
r=\aq\frac{-2Q^2}{5a(R+12c_2)}\cq^{1/5}\segue
 F(R)=\frac{Q^2}2\,\aq\frac{-5a(R+12c_2)}{2Q^2}\cq^{4/5}\,.
\label{deS3}\eeq
The latter function satisfies (\ref{due2}) and
a direct computation  using (\ref{fieldEq}) shows that the corresponding action
\beq
I=\int_{\cal M}\,\sqrt{-g}\,d^4x\,\aq \frac{Q^2}2\,
   \aq\frac{-5a(R+12c_2)}{2Q^2}\cq^{4/5}-\frac14\,F^{ij}F_{ij}\cq\,,\label{es2}
\eeq
gives the required solution
\begin{equation}
 B(r)=\frac{c_1}{r^2}+c_2r^2+\frac{Q^2}{5ar^3}\,.\label{es2B}
\end{equation}
In the third case here considered, that is $a\neq0,b\neq0$, the method does not work
owing to the presence of logarithmic terms which  prevent the explicit reconstruction
of the Lagrangian $F(R)$, which formally reads
\beq
F(R)=\frac12\,bR+\frac{k(b+4ar)}{2r^2}-\frac{c_0b^5}{2a^2r^2(ar+b)^2}\,.\label{GGG}
\eeq
However, if we choose $c_0=0$ then
the logarithm disappears from all equations and we get
\beq
a\neq0,b\neq0\,,\hs c_0=0
\segue
\ag\begin{array}{l}
B(r)=\frac{k}2+\frac{bk}{3ar}+c_2\,r^2+\frac{Q^2}{4br^2}\,,\\
R=-12c_2+\frac{k}{r^2}\,,\\
F(R)=\frac12\,bR+\frac{k(b+4ar)}{2r^2}\,.
\end{array}\cp
\eeq
These equations can be easily solved to get
\beq
r=\sqrt{\frac{k}{R+12c_2}}\segue F(R)=b(R+6c_2)+2a\,\sqrt{k(R+12c_2)}\,.\label{modellone}
\nn\eeq
This is a topological charged generalization  of the result found in Ref.~\cite{SSSsolutions}.
Also in this case one can directly verify that the field equations in (\ref{fieldEq}) are
satisfied.

\subsection*{Lifshitz like solutions}
Now we are going to consider the second case, that is $\al(r)\sim\log(r^z)$,
$z$ being an (in principle) arbitrary parameter.
With this special but important choice, Eq.~(\ref{two}) can be explicitly
solved, to get
\beq
f(R)=a r^{\frac12\at1+z+p(z)\ct}+b r^{\frac12\at1+z-p(z)\ct}\,,\hs\hs
p(z)=\sqrt{z^2+10z+1}\,,
\label{F1sol}\eeq
$a,b$ being arbitrary integration constants.
In the absence of an electric charge, that is $Q=0$, and choosing $b=0$ (or $a=0$)
also the Eq.~(\ref{tre}) can be explicitly solved, obtaining in this way
a complicated  expression for $B(r)$, valid for any $z$. It depends on two
arbitrary integration constants, say $c_{1,2}$,  and for $z\neq1$ it has the form
\beq
B(r)=\ga_0\,k+\frac{c_1}{r^{\ga_+}}+\frac{c_2}{r^{\ga_-}}\,,
\label{BGen}\eeq
$\ga_0,\ga_\pm$ being given function of $z$, which read
\begin{eqnarray}
a\neq0,b=0\segue
\ag\begin{array}{lll}
\ga_0&=& \frac{2}{(z-1)[p(z)-3(z+1)]}\,,\\
\ga_\pm&=& r^{\frac{7z-p(z)-1}{4}
        \pm\frac{i}2\,\sqrt{\frac{z^3+[p(z)-3]\,z^2-[8p(z)-69]\,z-23p(z)+41}{p(z)-3(z+1)}}}\,,
\end{array}\cp\end{eqnarray}
\begin{eqnarray}
a=0,b\neq0\segue
\ag\begin{array}{lll}
\ga_0&=& -\frac{2}{(z-1)[p(z)+3(z+1)}\,,\\
\ga_\pm&=& r^{\frac{7z+p(z)-1}{4}
    \pm\frac{1}2\,\sqrt{\frac{z^3-[p(z)+3]\,z^2+[8p(z)+69]\,z-23p(z)+41}
                         {p(z)+3(z+1)}}}\,.
\end{array}\cp\end{eqnarray}
The corresponding metric can be considered as a generalization of the one of
Clifton-Barrow (see below).
In the special case $z=1$, a logarithmic term is also present.

For example, for the following choices of the parameters, we get
\begin{eqnarray}
1)\:\:\:a\neq0,b=0,z=1/2\segue
\ag\begin{array}{lll}
B(r)&=&\frac{4k}{7}+\frac{c_1}{r^{7/2}}+c_2r\,,\\
R&=&-\frac{9c_2}{r}\,,\\
F(R)&\sim& 6k-\frac{81c_2^2}{R}\,,
\end{array}\cp
\end{eqnarray}
\begin{eqnarray}
2)\:\:\:a\neq0,b=0,z=2\segue
\ag\begin{array}{lll}
B(r)&=&-\frac{k}{7}+\frac{c_1}{r^7}+\frac{c_2}{r^2}\,,\\
R&=& \frac{4k}{r^2}\,,\\
F(R)&\sim& 30c_2-\frac{16k^2}{R}\,,
\end{array}\cp
\end{eqnarray}
\begin{eqnarray}
3)\:\:\:a\neq0,b=0,z=1\segue
\ag\begin{array}{lll}
B(r)&=&-\frac{c_1r^{-(3+\sqrt{3})}}{3+\sqrt{3}}-\frac{2 k\log\left(r\right)}{3+\sqrt{3}}+c_2\,,\\
R&=& \frac{12k\log r+2(9+\sqrt{3})k-6(3+\sqrt3)c_2}{(3+\sqrt3)\,r^2}\,,\\
F(R)&\sim&\left[\frac{R}{r}-\frac{4k}{r^3}
     +\frac{(54+30\sqrt3)c_2-12k\log r^{2+\sqrt3}}{(3+\sqrt{3})\,r^3}\right]\,r^{2+\sqrt{3}}\,,
\end{array}\cp
\end{eqnarray}
\begin{eqnarray}
4)\:\:\:a=0,b\neq0,z=1/2\segue
\ag\begin{array}{lll}
B(r)&=&2k+\frac{c_1}{r}+rc_2\,,\\
R&=& -\frac{5 k}{r^2}-\frac{9 c_2}{r}\,,\\
F(R)&\sim& R^{5/4}\,,\hs\hs\mbox{ if }c_2=0\,,\\
F(R)&\sim& R^{3/2}\,,\hs\hs\mbox{ if }k=0\,,\\
\end{array}\cp
\end{eqnarray}
\begin{eqnarray}
5)\:\:\:a=0,b\neq0,z=2\segue
\ag\begin{array}{lll}
B(r)&=&-\frac{k}{2}+\frac{c_1}{r^2}\,,\\
R&=&\frac{9 k}{r^2}\,,\\
F(R)&\sim&R^{3/2}\,,
\end{array}\cp\end{eqnarray}
\begin{eqnarray}
6)\:\:\:a=0,b\neq0,z=1\segue
\ag\begin{array}{lll}
B(r)&=&\frac{c_1 r^{-3+\sqrt{3}}}{-3+\sqrt{3}}+\frac{2 k\log\left(r\right)}{-3+\sqrt{3}}+c_2\,,\\
R&=&\frac{12k\log r+2(9-\sqrt3)k
     -6\left(3-\sqrt3\right)c_2}{(3-\sqrt3)\,r^2}\,,\\
F(R)&\sim&\left[\frac{R}{r}-\frac{4k}{r^3}
     +\frac{(54-30\sqrt3)c_2-12k\log r^{2-\sqrt3}}{(3-\sqrt{3})\,r^3}\right]\,r^{2-\sqrt{3}}\,,
\end{array}\cp\end{eqnarray}
The function $F(R)$ for the two examples with $z=1$ has not been explicitly
written in terms of the Ricci scalar, because it is a quite complicated expression
which we do not will use in what follows, while in the example (4)
we have also set $c_2=0$ (or $k=0$), in order to satisfy Eq.~(\ref{two}).
We observe that example (5) and the first part of example (4)
are particular cases  of the Clifton-Barrow metric,
corresponding to $\de=1/2$ and  $\de=1/4$, respectively.

In the previous section we have seen that a given metric can be the solution of different
actions. In the same way, a given action can give rise to different SSS-metrics.
For example, $F(R)\sim R^{3/2}$ has the Schwarzschild solution, but also the solutions
in the examples (4) and (5) above.

\subsection*{Topological Clifton-Barrow solutions}

Here we present  the topological generalization (without charge) of the Clifton-Barrow solution~\cite{CB},
which is a SSS metric of the form considered in the previous section, with
\begin{eqnarray}
\alpha(r)&=&\log\left[\left(\frac{r}{r_0}\right)^{\delta(1+2\delta)/(1-\delta)}\left(\frac{(1-2\delta+4\delta^2)(1-2\delta-2\delta^2)}{(1-\delta)^2}\right)^{1/2}\right]\,,\label{CBalpha}\\ \nonumber\\
B(r)&=&\frac{(1-\delta)^2}{(1-2\delta+4\delta^2)(1-2\delta-2\delta^2)}\left(k-\frac{C}{r^{(1-2\delta+4\delta^2)/(1-\delta)}}\right)\,,
\label{CBB}
\end{eqnarray}
\phantom{line}\\
$C$ being an integration constant and $\delta\neq1$ an arbitrary parameter.

The Ricci scalar reads
\begin{equation}
R=\frac{6\delta(1+\delta)}{(2\delta^2+2\delta-1)}\left(\frac{1}{r^2}\right)\,,
\label{CBR}
\end{equation}
\phantom{line}\\
and, for $\de\neq-1$, on the solution we obtain
\begin{equation}
F(R)\sim R^{1+\de}\,,
\label{CB}\end{equation}
One can directly verify that (\ref{CBB}) are solutions of the field equations
corresponding to the Clifton-Barrow Lagrangian (\ref{CB}).
The Clifton-Barrow metric is a particular case of the one considered in the previous
section, Eq.~(\ref{BGen}), with one of the two constants ($c_1$ or $c_2$) dropped off and, in fact,
it can be obtained using the method described above.

\section{Charged topological SSS solutions in Weyl conformal gravity}
\label{S:Weyl}

In this Section we generalize  the higher gravity black hole solution
of Riegert \cite{r,m} and its topological version~\cite{klem} in the Weyl model
including the charge.
We also present general solutions of pure Weyl gravity derived with the method
of  conformal transformations.

To start, we write down the conformal invariant action of the model, in the form
\beq
I=I_W+I_{EM}\,,
\eeq
where the gravitational part is given by Weyl's conformal gravity
\begin{equation}
I_W=3 w \int_{\mathcal M} d^4 x\,\sqrt{-g}\,\,C^2\,,
\label{actionW}\end{equation}
where $w$  is a dimensionless parameter, usually restricted to the values  $w>0$, and
$C^2=C_{ijrs} C^{ijrs}$  is the square of the Weyl tensor $C_{ijrs}$.
It is a scalar quantity which, in $4-$dimensions, is given by
\begin{equation}
C^2=\frac{1}{3}R^{2}-2R_{ij}R^{ij}+R_{ijrs}R^{ijrs}\,.\label{Weylsquare}
\end{equation}
The  conformal gravity model has a very interesting future domain and its phenomenology
has been investigated in Ref.~\cite{m2}.

Again, we look for SSS solutions of the form in (\ref{metric0}). With this choice, one has
$C^2=A(r)^2/3r^4$, where
\beq
A(r)=r^2B''+\left(3r\al'-2\right)rB'+2\left[r^2\al''+r^2(\alpha')^2-r\alpha'+1\right]\,B-2 k\,.
\label{A(r)}\end{eqnarray}
The total action reads now
\begin{equation}
 I=V_k\int dt \int dr\,e^{\al(r)}\left[\frac{wA(r)^2}{r^2}+\frac{Q^2}{2r^2}\right]\,.
\end{equation}
We are dealing  with a higher-order Lagrangian system, because  the Lagrangian depends
on the first and second derivatives of the unknown functions  $\al(r)$ and $B(r)$.
The corresponding equations of motion (after simplification) read
\begin{eqnarray}
wA''B+wA'\at\frac{B}{r}+\frac{B'}2\ct-\frac1{4r^2}\,\aq w(A^2+4kA)-\frac{Q^2}{2}\cq=0\,,
\label{bw}\end{eqnarray}
\begin{eqnarray}
A''+A'\at\frac2r-\al'\ct=0\,.
\label{aw}\end{eqnarray}
In order to find explicit solutions, we proceed as in the previous section,
first considering the case $\al(r)=$constant. With this choice, one immediately gets
\beq
A(r)=6c_0+\frac{6b_1}{r}\,,
\nn\eeq
and Eqs.~(\ref{A(r)}) and (\ref{bw}) give the solution
\begin{equation}
B(r)=k+3c_0+\frac{b_1}{r}+c_1\,r+c_2r^2\,,\hs\hs
             b_1\neq0\,,\hs c_1=\frac{c_0(2k+3c_0)}{b_1}-\frac{Q^2}{24wb_1}\,,
\label{B(r)}
\end{equation}
$c_0,b_1,c_2$ being arbitrary integration constants.
The metric here obtained is the generalization to the charged topological case
of the black hole Riegert solution~\cite{r,klem}.

The case $b_1=0$ has to be treated separately, because the limit in (\ref{B(r)}) is singular.
In this case, we have
\beq
B(r)=w\sqrt{k^2+\frac{Q^2}{8w}}+c_1\,r+c_2r^2\,,\hs\hs b_1=0\,,\hs\hs w=\pm1\,,
\eeq
$c_1,c_2$ being arbitrary integration constants.
By conveniently choosing the integration constants of the Riegert solution one obtains the Schwarzschild,
Schwarzschild-de Sitter(AdS) or the de Sitter metric, correspondingly.

In principle, one can try to find other solutions by solving the equations for
 a non constant $\al(r)$, but they are really complicated and so
it is convenient to proceed in a different way, by using the fact that the
Weyl-Maxwell action is conformally invariant.
This implies that, if $ds^2$
is a solution of the field equations, then also
$d\tilde s^2=\Omega^2(x^k)ds^2$
is a solution, $\Omega$ being an arbitrary (smooth) function of the coordinates.
In particular, starting from a SSS solution of the form (\ref{metric0}) and choosing
$\Om=\Om(r)>0$ to depend on the radial coordinate only, one obtains again
a static and spherically symmetric solution, which can be
set in the form  (\ref{metric0}), with the change of radial variable
$\tilde r= r\,\Om(r)$. In fact,
\beq
d\tilde s^2=-\Om^2(r)e^{2\al(r)}B(r)\,dt^2+\frac{\Om^2(r)\,dr^2}{B(r)}+
r^2\Om^2(r)\,d\si^2_k\,,
\label{ConfTrans}\eeq
and this assumes the form
\beq
d\tilde s^2=-e^{2\tilde\al(r)}\tilde B(r)\,dt^2+\frac{dr^2}{\tilde B(r)}+r^2\,d\si^2_k\,,
\eeq
where, for simplicity, we have re-stated the original variable $r$ by
$r\Om(r)\to r$, $\Om(r)\to r/\Xi(r)$ and
\beq
\tilde B(r)=B(\Xi(r))\aq\frac{\Xi(r)}{r\Xi'(r)}\cq^2\,,\hs\hs
e^{2\tilde\al(r)}\,\tilde B(r)=e^{2\al(\Xi(r))}\,B(\Xi(r))\,\aq\frac{r}{\Xi(r)}\cq^2\,.
\eeq
In this way we have obtained a class of SSS solutions for the action (\ref{actionW})
in the form (\ref{metric0}), specified by the functions $\al(r),B(r)$.
Any solution in this class is related to the original one
by means of an arbitrary, positive (smooth) function $\Xi(r)$.
As one can immediately see, $\Xi(r)=r$ corresponds to the unitary transformation $\Om(r)=1$.
On the contrary, starting from Riegert's solution (\ref{B(r)}) and choosing
an arbitrary $\Xi(r)\neq r$ one obtains a class of solutions with
$\al(r)\neq0$ given by means of the equation
\beq
e^{2\tilde\al(r)}=\aq\frac{r}{\Xi(r)}\cq^4\,\aq\Xi'(r)\cq^2\segue
\frac{d}{dr}\aq\frac1{\Xi(r)}\cq=-\frac{e^{\tilde\al(r)}}{r^2}\,.
\label{alRiegert}\eeq
from which we see that, for fixed $\tilde\al(r)$,
one can in principle derive the function $\Xi(r)$ which provides the desired transformation.

\subsection*{Riegert-Lifshitz solutions}\label{S:RLS}
This is a class of solutions similar to the ones obtained for the $F(R)$ case,
but starting from that of Riegert, performing a conformal transformation as described in the
previous section, and imposing $\tilde\al(r)$ to be of the form
\begin{equation}
\tilde{\alpha}(r)=\log(\gamma\tilde{r}^z)\,,\hs\hs\ga>0\,,\label{B22}
\end{equation}
where $\gamma$ is a dimensional constant and $z$ an arbitrary parameter.

From (\ref{alRiegert}), we immediately get
\beq
\Xi(r)&=&-\frac{1}{\ga\log(qr)}\,,\hs\hs z=1\,,
\nn\\
\Xi(r)&=&-\frac{1}{q+\ga r^{z-1}/(z-1)}\,,\hs\hs z\neq1\,,
\nn\eeq
$q$ being an integration constant.

In the first case ($z=1$) for $\tilde B(r)$ we easily obtain
\beq
\tilde B(r)=\frac{c_1}{\ga^2}-\frac{c_1}\ga\,\log(q r)
+(k+3c_0)\,\log^2(q r)-\ga b_1\,\log^3(q r)\,,
\eeq
while, in the second case ($z\neq1$) and for simplicity choosing $q=0$,
for $\tilde B(r)$ we get
\beq
\tilde B(r)=\tilde c_0+\tilde b_1\,\frac{r^z}{r}
    +\tilde c_1\,\frac{r}{r^z}+\tilde c_2\,\frac{r^2}{r^{2z}}\,,\hs\hs
\ag\begin{array}{l}
\tilde b_1=\frac{\ga b_1}{(z-1)^3}\,,\\
\tilde c_0=\frac{k+3c_0}{(z-1)^2}\,,\\
\tilde c_1=-\frac{c_1}{\ga(z-1)}\,,\\
\tilde c_2=\frac{c_2}{\ga^2}\,,
\end{array}\cp
\hs\hs z\neq1\,.
\label{tildeCoeff}\eeq



 \section{Black hole solutions and event horizons}

The SSS metric may actually describe a black hole. We recall that an event horizon exists as soon as there exists a positive solution $r_H$ of
\begin{equation}
B(r_H)=0\,,\quad\quad B'(r_H) \neq 0\,. \label{BHconditions}
\end{equation}
The second condition describes a non extremal BH.

For example, in the well known case of constant curvature, namely (\ref{deS1})
with $Q=0$ and  $k=0$ or $k=-1$,
the conditions (\ref{BHconditions}) are satisfied if both
$c_1$ and the cosmological constant $\Lambda$ are negative (AdS);
on the contrary, if $k=1$, one has a black hole solution
whatever the sign of $\Lambda$ if  $c_1<0$ again. A further restriction may be $B'(r_H)>0$, namely the Killing surface gravity at the horizon to be positive.

For the topological Clifton-Barrow solution (\ref{CBB}), one explicitly finds
\begin{equation}
r_{H}=(C/k)^{(1-\delta)/(1-2\delta+4\delta^2)}\,,\label{CBr}
\end{equation}
which is positive if $C/k>0$. In order to have a black hole, one has also to impose
$(1-2\delta+4\delta)/(1-\delta)>0$.

Finally, in the case of the charged topological Riegert solution (\ref{B(r)}),
requiring positive surface gravity one has
\begin{equation}
k+3c_0+c_1\,r_H+c_2 r_H^2+\frac{b_1}{r_H}=0\,,\hs\hs
c_1\,+2c_2 r_H-\frac{b_1}{r_H^2}>0\,,\hs\hs r_H>0\,.
\end{equation}
If $b_1<0$ and $c_2>0$, then
it is easy to show that there always exists a positive root of $B(r_H)=0$,
independently of the values of $c_0$ and $c_2$.

In the following we shall assume to deal with (topological)
SSS black hole solutions satisfying conditions (\ref{BHconditions}), with the additional requirement $B'(r_H)>0$.

\section{The Wald entropy}
\label{S:WE}

With regard to the evaluation of the entropy associated with a black holes solution, there exists a general
method due to Wald \cite{Wald}, see also \cite{Visser:1993nu, FaraoniEntropy}. The explicit calculation of the black
hole entropy $S_W$ is provided by the formula
\begin{equation}
S_W=-2\pi\oint_{\begin{array}{c}r=r_H \\ t = \mbox{const}\end{array}}
\left. \left(\frac{\delta \mathcal L}{\delta R_{ijrs}}\right)\right|_{r=r_H}\,
       e_{ij} e_{rs}\sqrt{h_2}\,d\rho\,d\phi\,.
\label{Wald}
\end{equation}
Here $\sqrt{h_2}\,d\rho\,d\phi=dV_k$ represents
the induced volume form on the bifurcate surface $r =r_H,t=\mathrm{const}$,
while $\mathcal L=\mathcal L(g_{ij},R_{ijrs},\nabla_kR_{ijrs},...) $
is the Lagrangian density of an arbitrary theory of gravity.
The antisymmetric variable $e_{ij}$
is the binormal vector to the (bifurcate) horizon.
It is normalized so that $e_{ij}e^{ij}=-2$ and, if the metric has the form in
(\ref{metric0}), turns out to be
\begin{equation}
\epsilon_{ij}=\sqrt{-g_{00}\,g_{11}}\,(\delta^0_{i}\delta^1_{j}-\delta^0_{j}\delta^1_{i})
=e^{\al(r)}\,(\delta^0_{i}\delta^1_{j}-\delta^0_{j}\delta^1_{i})\,,
\end{equation}
$\delta^i_j$ being the Kronecker symbol. As a consequence, one has
\beq
S_W=-8\pi\mathcal{A}_H\,e^{2\alpha(r_H)}\,
           \left(\frac{\delta\mathcal L}{\delta R_{0101}}\right)\Big\vert_{r=r_H}
\label{waldbis}\,.
\eeq
$\mathcal A_H=V_k r_H^2$ being the area of the horizon.
Note that the expression in (\ref{waldbis})
is valid only because we are considering an SSS metric of the form (\ref{metric0}). Furthermore, the
 variation of the Lagrangian density in (\ref{Wald}) and (\ref{waldbis}) has to be  performed
by considering $R_{ijrs}$ and $g_{ij}$ as independent quantities.
As an example,
\begin{equation}
\frac{\delta R}{\delta R_{ijrs}}=
\frac{1}{2}\left(g^{ir}g^{js}-g^{is}g^{jr}\right)\segue
   \frac{\delta R}{\delta R_{0101}}=g^{00}g^{11}=-e^{-2\al(r)}\,.
\end{equation}
For $F(R)$-modified gravity the Lagrangian density has the form
$\mathcal L=F(R)+\mathcal L_{EM}$ and, since the electromagnetic
Lagrangian does not depend on the Riemann tensor, one immediately obtains
\begin{equation}
S_W=\frac{{\cal A}_H\,f(r_H)}{4}\,,\hs\hs
     f(r_H)=\left.\frac{dF(R)}{dR}\right|_{r=r_H}\,.
\label{waldF(R)}\end{equation}
For pure Einstein gravity $\al=0,f(r_H)=1$ and one recovers the well known Area Law.

\subsection{Wald's entropy for conformally invariant gravity models}

In general the Wald entropy is not invariant with respect to conformal transformations,
but when the action is conformally invariant, then also  the entropy does not change.
This can be easily seen by recalling that,
under a conformal transformation as in (\ref{ConfTrans}), one has
\beq
\tilde g_{ij}=\Om^2g_{ij}\,,\hs\hs
\tilde g^{ij}=\Om^{-2}g^{ij}\,,\hs
\sqrt{-\tilde g}=\Om^4\,\sqrt{-g}\,,
\eeq
\beq
\tilde R_{ijrs}=\Om^2\,R_{ijrs}+U(g,\nabla\Om)\,,\hs\hs
   \tilde{\mathcal L}=\mathcal L(\tilde R_{mnpq},...)=\Om^{-4}\mathcal L\,,
\eeq
where $U$ is a function which does not depend on the Riemann tensor.
If the action is invariant, both metrics $g_{ij}$ and
$\tilde g_{ij}$ are black hole solutions and thus, for the metric in (\ref{metric0}),
we get
\beq
\tilde S_W&=&-2\pi\,\oint_{\begin{array}{c}r=r_H\\t=\mbox{const}\end{array}}\left.
              \left(\sqrt{\tilde g_{00}\tilde g_{11}\tilde g}\,\,
               \frac{\delta\tilde{\mathcal L}}{\delta\tilde R_{0101}}\right)\,
                \right|_{r=\tilde r_H}\,d\rho\,d\phi
\nn\\
  &=&-2\pi\,\oint_{\begin{array}{c}r=r_H\\t=\mbox{const}\end{array}}\left.
           \left(\Om^2\sqrt{g_{00}g_{11}g}\,\,
            \frac{\delta{\mathcal L}}{\delta R_{0101}}\,
             \frac{\delta R_{0101}}{\delta\tilde R_{0101}}\right)\,
              \right|_{r=r_H}\,d\rho\,d\phi
\nn\\
   &=&-2\pi\,\oint_{\begin{array}{c}r=r_H\\t=\mbox{const}\end{array}}\left.
         \left(\sqrt{-g}\,\,\frac{\delta\mathcal L}{\delta R_{0101}}\right)\,
          \right|_{r=r_H}\,d\rho\,d\phi=S_W\,.
\eeq
As it is well known, also the Killing surface gravity is a conformally invariant quantity and
this means that, for example, all black holes described in Sect.~(\ref{S:RLS})
have the same entropy and surface gravity.

As an example, let us consider Weyl's conformal gravity. From  (\ref{actionW}), one has  $\mathcal L=3wC^2+\mathcal L_{EM}$. Making use of  (\ref{Weylsquare}) and (\ref{waldbis}), we get
\beq
S_W&=&-24\pi w\,\mathcal A_He^{2\al(r_H)}\,
   \aq2R^{0101}-(g^{00}R^{11}+g^{11}R^{00})+\frac13\,g^{00}g^{11}R\cq_{r=r_H}
 \nn\\
    &=&-24\pi w\,\mathcal A_H\,\aq\frac{1}{3}B''+\at\alpha'-\frac{2}{3r}\ct\,B'
               -\frac{2k}{3r^2}\cq_{r=r_H}\,.
\label{Waldgen}
\eeq
In the particular case of Riegert's solution (\ref{B(r)}), the latter equation
simplifies to
\begin{equation}
S_W=-48\pi wV_k\left(\frac{b_1}{r_H}+c_0\right)\,.\label{Entropy}
\end{equation}

\section{First Law for Black Hole solutions}

In Ref.~\cite{SSSEnergy} an expression for the uncharged BH energy in a class of $F(R)$ modified gravities has been proposed and identified with a quantity proportional to the unique  constant of integration, which appears in the explicit solutions. The identification is achieved by making use of the
derivation of the First Law of black hole thermodynamics from the equations of motion of $F(R)$-gravity,
evaluating independently the entropy via Wald's method and the
Hawking temperature~\cite{HT} via quantum mechanical methods in curved space-times.
In this section we will briefly revisit the case of $F(R)$-gravity by discussing the results of Ref.~\cite{SSSEnergy} to the topological Maxwell solutions. In the following, for the sake of simplicity we shall assume the charge $Q$ as being a given quantity and not subject to thermodynamical variations.  This, of course, is a restriction, implying that in the First Law the
thermodynamical potential associated with the variation of the charge will be absent.

From the equation of motion (\ref{one}) of $F(R)$-gravity evaluated on the event horizon
(namely $B(r_H)=0$) and recalling the expression (\ref{waldF(R)}) for the Wald entropy,
we easily get
\begin{equation}
B'(r_H)\frac{\partial S_W}{\partial r_H}
     =\frac{V_k}4\,\left[(F-Rf)r_H^2+2kf\right]_{r=r_H}+\frac{V_k\,Q^2}{8r_H^2}\,.
\label{EquazionePrincipe}\end{equation}

It is well known \cite{HT} that all black holes have a characteristic
temperature $T_K$ related to the existence of an  event horizon and, with the associated entropy,  they have  thermodynamical
properties. For the static black hole described by the metric in (\ref{metric0}),
the Killing-Hawking temperature is proportional to the Killing surface gravity and has the explicit form
\begin{equation}
T_K=\frac{e^{\alpha(r_H)}B'(r_H)}{4\pi}\,,\label{HT}\,.
\end{equation}
This is a conformally invariant quantity (see for example \cite{jac,nada}. This make sense if the conformally related metric is a black hole solution with event horizon at $r=r_H$, and with regard to this issue a simple proof of conformal invariance of the BH temperature is presented in the Appendix. The above relation between surface gravity and BH temperature  is a very robust and well known result (see for example \cite{VisserHaw}).
The expression (\ref{HT}) has been derived in several different ways, like the original Hawking derivation,
by eliminating the conical singularity in the corresponding Euclidean metric
or either by making use of tunneling methods recently introduced in Refs.~\cite{PW,Nadalini}
and discussed in details in several papers.

As in general relativity,  the temperature appears in a natural way in  (\ref{EquazionePrincipe}).
Furthermore, if the entropy depends only on $r_H$, as it happens in a large class of  $F(R)$ theories,
Eq.~(\ref{EquazionePrincipe}) can be used to derive  the First Law of black holes thermodynamics
and employed to define a specific BH  energy~\cite{SSSEnergy}.
Thus, we may write
\begin{equation}
\frac{V_ke^{\al(r_H)}}{16\pi}\,
         \left[(F-Rf)r_H^2+2kf\right]_{r=r_H}\,dr_H
       =T_{K}dS_W-pd{\cal V}:=dE_{K}\,,
\label{differentialform}\end{equation}
\beq
{\cal V}=\frac13\,V_kr_H^3\,,\hs\hs p=\mathrm{e}^{\alpha(r_H)}\frac{Q^2}{32\pi\,r_H^4}\,,
\eeq
where ${\cal V}$ is the volume of the black hole and
$p$ may be interpreted as the work  term associated with  the  electromagnetic field.

If the differential of the function in (\ref{differentialform})
is interpreted  as  variation  due to an infinitesimal
change of the size $r_H$ of the black hole, one may write
\begin{equation}
 E_{K}:=\frac{V_k}{16\pi}\int\,dr_H\,e^{\alpha(r_H)}
         \left[(F-Rf)r_H^2+2kf\right]\Big\vert_H
\,.\label{BHEnergy0}
\end{equation}
We recall that this is valid only if $r_H$ depends on a unique
variable parameter, which will be proportional to  the mass of the black hole.
If $r_H$ and, as a consequence, also the entropy depend on other variables,
then other thermodynamics potentials will appear and the expression for the energy
cannot be computed by means of the above expression.

In what follows we shall apply these considerations to some exact solutions of $F(R)$-gravity.
We will see that, in some interesting cases, Eq.~(\ref{BHEnergy0})
allows to identify the energy of the black hole with a suitable integration (mass) constant
which appears in the explicit solution.

As we have seen in (\ref{Bab}), the function $B(r)$ in general depends on three or four
parameters, but in some cases some of them explicitly appear in the
``reconstructed'' Lagrangian and thus we can assume them to be a fixed constant
of the model. Only the remaining parameters (the integrating constant of the solution),
have to be varied in order to obtain thermodynamical equations.
If only one free parameter survives, then the corresponding variation can be related
to the energy via the first law of black holes, as explained above.

\subsection*{BH energy in $F(R)$ gravity: Examples}

Here we explicitly compute the energy for some models we have derived in Sect.\ref{S:SSSsol}.
To start with~, let us consider the first model in (\ref{Bab}), described by (\ref{deS1}).
Since $F(R)=(R-2\La)$ and $Q$ is kept fixed, the only free parameter is $c_1$, and from (\ref{waldF(R)}),
(\ref{HT}), and (\ref{BHEnergy0}), we get
\beq
S_W=\frac{A_H}{4\pi}\,,  \hs
T_K=\frac{k-\La r_H^2-Q^2/4r_H^2}{4\pi r_H}\,,\hs
E_K=\frac{V_k}{8\pi}r_H\left(k-\frac{\Lambda}{3}r_H^2\right)=\frac{V_k}{8\pi}\,c_1-\frac{V_k Q^2}{32\pi r_H} \,,
\label{STE1}\eeq
where we have used $B_H=0$, namely
\beq
k-\frac{c_1}{r_H}-\frac{\Lambda}{3}r_H^2+\frac{Q^2}{4 r_H^2}=0\,.
\eeq
Furthermore, for $\Lambda=0$, one has  $k=1$, and $v_k=4\pi$, $E_k=\frac{r_H}{2}$. Thus
\beq
c_1=r_H+\frac{Q^2}{4r_H}\,.
\eeq
This may be interpreted as  the Christodoulou-Ruffini irreducible mass relation \cite{remo}, which allows to identify the BH mass $M$ as $M=2c_1$


As a second example let us consider the non trivial model in (\ref{mod2})
in the two special cases $(Q=0,k\neq)0$ and $(Q\neq0,k=0)$.
These are described by the actions in (\ref{es}) and in (\ref{es2}), respectively,
and the energy can be computed by the method described above.

In the first case, entropy, temperature
and energy, respectively, follows from (\ref{waldF(R)}), (\ref{HT}), and (\ref{BHEnergy0}).
We obtain,
\beq
S_W=\frac{A_H}{4\pi}\,ak\,r_H\,,  \hs\hs
T_K=-\frac1{4\pi}\,\at\frac{k}{r_H}+\frac{4c_1}{r_H^3}\ct\,, \hs\hs
E_K=\frac{3aV_0}{32\pi}\,c_1\,.
\eeq
We see that the entropy and energy are positive only if all parameters $a,k,c_1$ have the same
sign, but this means they must be all negative, in order to have a positive temperature, in particular $k=-1$.
On the contrary, the given constant $c_2$ has to be positive for having $r_H>0$, yielding an asymptotically AdS space-time.


The second case in quite similar to the previous one. Now $F(R)$ is given in (\ref{deS3}),
and we get
\beq
S_W=-\frac{A_H}{4\pi}\,a\,r_H\,,  \hs\hs
T_K=\frac1{4\pi}\,\at\frac{-Q^2}{ar_H^4}-\frac{4c_1}{r_H^3}\ct\,, \hs\hs
E_K=-\frac{3aV_k}{16\pi}\,c_1-\frac{V_kQ^2}{32\pi r_H}\,.
\eeq
Here we recall that $a <0$.

As a further example we consider the topological Clifton-Barrow solutions
(\ref{CBalpha})-(\ref{CBB}) of model $F(R)\sim R^{\delta+1}$.
One easily finds
\begin{equation}
E_K=\frac{\Psi_\delta}{r_0^{\delta(1+2\delta)/(1-\delta)}}\,\left(\frac{C}{k}\right)\,,
\label{CBEn}\end{equation}
where we have introduced the dimensionless constant
\begin{equation}
\Psi_\delta=\left(\frac{2^{\delta-1}3^{\delta}\delta^{\delta}(\delta-1)^2(\delta+1)^{\delta
+1}}{\sqrt{1-2\delta-2\delta^2}\sqrt{1-2\delta+4\delta^2}}\frac{1}{(2\delta^2+2\delta
-1)^{\delta}}\right)\,.\end{equation}
In order to have the BH horizon and a positive Killing temperature,
the range of the parameter $\delta$ in the latter equation has to be restricted
to the values already discussed.
Some additional restrictions are also necessary to get a positive entropy.
With such restrictions the energy in (\ref{CBEn}) is well defined and positive.
As expected, in the limit  $\delta\rightarrow 0$ it reduces to the
Misner-Sharp mass of general relativity.

Let us pause to summarize these noticeable results. In Eq.~(\ref{differentialform})
we have extended the validity of the Killing/Hawking temperature $T_K$,
which expression is derived in general relativity by using quantum mechanics,
to topological Maxwell-$F(R)$-gravity.
We stress that the Killing/Hawking temperature is well defined by the metric.
In this way, we identify the black hole energy making use of the derivation
of the first law of black hole thermodynamics from the equations of motion,
by evaluating the related black hole entropy via Wald's method,
and the Killing/Hawking temperature as the semiclassical result stemming from Hawking's radiation,
as in the case of general relativity.

\subsection{Black hole Energy in Maxwell-Weyl conformal gravity}

We have seen in Sect.~\ref{S:Weyl} that the Lagrangian of Einstein-Weyl's
gravity contains only a dimensionless parameter $w$, while the solution $B(r)$
depends on three arbitrary integration constants $c_0,b_1,c_2$. In this case the  $r_H$ coordinate of the black
hole horizon will depend on several integration constants. In Ref.~\cite{SSSEnergy}, a particular case with $c_0$ and $c_2$ held fixed has been considered.

In the general case, when asymptotically the solution is anti de Sitter,
the energy of the BH may be defined by means of the Euclidean action
and the First Law of black holes thermodynamics holds with additional
thermodynamical potentials (see Ref.~\cite{Pope} for more details). With regard to this issue, let us imagine we are dealing with a BH solution, with $r_H=r_H(c_n)$ being the larger positive solution of $B(r_H)=0$, and $c_n$ 
constants of integration. Then, one has
\beq
0=dB=\sum_n\at\frac{\partial B}{\partial c_n}\,dc_n
         +B'(r_H)\,\frac{\partial r_H}{\partial c_n}\,dc_n\ct\segue
B'(r_H)\,\frac{\partial r_H}{\partial c_n}
=-\frac{\partial B}{\partial c_n}\,dc_n\,,
\eeq
where the $c_n$ are all independent and  $B'(r_H)$ is the partial derivative with respect to $r$
evaluated on the horizon. Recall that  $T_K$ and $S_W$ are computable, for example, from (\ref{Entropy})
\beq
S_W=-48\pi w V_k\,\at\frac{b_1}{r_H}+c_0\ct\,.
\eeq
Thus,  we get
\beq
T_KdS_W=12V_k\,(c_1\,dc_0-4c_2db_1-2b_1dc_2)+\frac{Q^2V_k}{2wb_1}\,dc_0\,,
        \hs\hs c_1=\frac{c_0(2k+3c_0)}{b_1}\,.
\eeq
In the particular case when  $c_0$ and $c_2$ are held fixed, one recovers the result of reference \cite{SSSEnergy}, namely
\beq
E_K=-48V_kc_2  b_1\,.
\eeq
In the general case,  all the constants of integration are true thermodynamic variables and, when it is possible, one has to make use  of Euclidean methods  in order to identify the energy. Then the First Law is shown to held true but with additional thermodynamic potentials, as explained in Ref.~\cite{Pope}.

\section{Conclusions}

In this paper, we have considered several example of extended models of gravity coupled with the electromagnetic field and we have studied charged topological static spherical black hole solutions, extending previous works.
New black hole solutions have been presented.  For modified  $F(R)$ gravity, we have investigated the First Law of BH thermodynamics, and
for a important class of topologically charged black hole solutions, and assuming the charge of the black hole as a given fixed quantity, we have obtained an $F(R)$ BH energy formula. In the particular case of general relativity, our expression for the  BH energy has been shown to be equivalent to a relation which defines the irreducible mass introduced by  Christodoulou and Ruffini.

We have also considered the Maxell-Conformal gravity model. As we have shown in Sect.~\ref{S:WE}, the entropy is conformally invariant for
Maxwell-Weyl's gravity and, since the temperature is conformally invariant too (see the Appendix),
we have
\beq
T_KdS_W=\tilde{T}_Kd\tilde{S}_W\,,
\nn\eeq
where the quantities on the right-hand side refer to the solution obtained by a conformal
transformation of the kind considered in \ref{S:Weyl}.
This means that, for example, all thermodynamical quantities related to the
Riegert-Lifshitz solutions considered in Sect.~\ref{S:RLS}
can be computed directly by using the general formulas or, more easily, by
replacing $c_n$ with $\tilde c_n$ coefficients
by means of the relations in (\ref{tildeCoeff}).

As a final remark, it is reasonable to check now in detail if the several examples of modified gravity models we have discussed in this paper do represent in fact feasible or viable models. For example, in the  $F(R)$ class, one should have \cite{de}
\beq
\frac{\partial F(R)}{\partial R}>0\,,\quad \frac{\partial^2 F(R)}{\partial^2 R}>0\,,
\eeq
for  $R > R_0$, $R_0$ being the current Ricci scalar curvature. In the case of the Clifton solution, these inequalities are satisfied as
soon as $\delta$ is positive and small. The same check may be also performed for the other classes of $F(R)$ models.

\section*{Appendix}
In this Appendix we give a simple proof of the conformal invariance of the Killing-Hawking temperature for an SSS BH space-time of the form
\begin{equation}
ds^2=-e^{2\alpha(r)}B(r)dt^2+\frac{d r^2}{B(r)}+r^2\,d\si^2_k\,.
\label{}\end{equation}
The event horizon is located at $B(r_H)=0$, and the related Killing-Hawking temperature is $T_K=e^{\alpha_H}B'_H$.

Let us perform a regular conformal transformation, with conformal factor $\Omega(r)$,
\begin{equation}
d \tilde{ s}^2=\Omega^2(r)d s^2=-e^{2\tilde{\alpha}(r)}\tilde{B}(r)dt^2+\frac{d r^2}{\tilde{B}(r)}+\tilde{r}^2\,d\si^2_k\,,
\label{metriccon}\end{equation}
where
\beq
\tilde{B}(r)=\frac{B(r)}{\Omega^2(r)}\,, \hs \hs e^{2\tilde{\alpha}(r)}=\Omega^4 e^{2\alpha(r)}\,.
\eeq
In the new metric, since $\tilde{r}=\Omega(r)r$, the event horizon is still located at $r=r_H$. As a consequence, the
new Killing-Hawking temperature is   $\tilde{T_K}=e^{\tilde{\alpha_H}}\tilde{B}'_H$. Now,
\beq
\tilde{B}'_H=\frac{B'_H}{\Omega^2_H}\,,
\eeq
and
\beq
e^{\tilde{\alpha}_H}=\Omega^2_H e^{\alpha_H}\,,
\eeq
and then it follows that
\beq
\tilde{T_K}=e^{\alpha_H}B'_H =T_K\,.
\eeq
\medskip

\noindent{\bf Acknowledgments.}
LS has been supported in this work by a Short Visit Grant, to Barcelona, of the European Science Foundation on New Trends and Applications of the Casimir Effect. EE was supported in part by MICINN (Spain), grant PR2011-0128 and project FIS2010-15640, by the CPAN Consolider Ingenio Project, and by AGAUR (Generalitat de Ca\-ta\-lu\-nya), contract 2009SGR-994, and his research was partly carried out while on leave at the Department of Physics and Astronomy, Dartmouth College, NH, USA.


{}


\begin{thebibliography}{}

\bibitem{review6}S.~Nojiri, S.~D.~Odintsov, {\em Int.\ J.\ Geom.\ Meth.\ Mod.\ Phys.}, {\em 4}, 115 (2007) arXiv: hep-th/0601213;
S.~Nojiri and S.~D.~Odintsov,
Phys.\ Rept.\  {\bf 505}, 59 (2011)
[arXiv:1011.0544 [gr-qc]].

\bibitem{review7}T.~P.~Sotiriou, V.~Faraoni, {\em Rev. Mod. Phys.}, {\em 82}, 451 (2010).

\bibitem{guido}G.~Cognola, E.~Elizalde, S.~Nojiri, S.~D.~Odintsov and S.~Zerbini, JCAP  {\bf 0502}, 010 (2005).

\bibitem{guido2}G.~Cognola and S.~Zerbini, J.\ Phys.\ A  {\bf 39}, 6245 (2006) [arXiv:hep-th/0511233].
G.~Cognola, L.~Sebastiani and S.~Zerbini, to appear in proceedings MG12, arXiv:1006.1586 [gr-qc] (2009).

\bibitem{Far}V.~Faraoni, Phys.\ Rev.\ D  {\bf 74}, 104017 (2006) [arXiv:astro-ph/0610734].

\bibitem{Monica}G.~Cognola, M.~Gastaldi and S.~Zerbini, Int.\ J.\ Theor.\ Phys.\  {\bf 47}, 898 (2008) [arXiv:gr-qc/0701138].

\bibitem{Saw}W.~Hu and I.~Sawicki, Phys.\ Rev.\ D  {\bf 76}, 064004 (2007) [arXiv:0705.1158 [astro-ph]].

\bibitem{seba10}
E.~Elizalde, S.~D.~Odintsov, L.~Sebastiani and S.~Zerbini,
Eur.\ Phys.\ J.\  C {\bf 72} (2012) 1843
[arXiv:1108.6184 [gr-qc]];
E.~Elizalde, S.~Nojiri, S.~D.~Odintsov, L.~Sebastiani and S.~Zerbini,
Phys.\ Rev.\  D {\bf 83} (2011) 086006
[arXiv:1012.2280 [hep-th]];
G.~Cognola, E.~Elizalde, S.~Nojiri, S.~D.~Odintsov, L.~Sebastiani and S.~Zerbini, Phys.\ Rev.\ D  {\bf 77}, 046009 (2008) [arXiv:0712.4017 [hep-th]];
G.~Cognola, E.~Elizalde, L.~Sebastiani and S.~Zerbini,
Phys.\ Rev.\  D {\bf 83} (2011) 063003
[arXiv:1007.4676 [hep-th]].

\bibitem{Od1} S.~Nojiri and S.~D.~Odintsov, Phys. Lett. B  {\bf 657}, 238 (2007) [arXiv:0707.1941 [hep-th]].

\bibitem{Od2}S.~Nojiri and S.~D.~Odintsov, Phys.\ Rev.\ D  {\bf 77}, 026007 (2008) [arXiv:0710.1738 [hep-th]].

\bibitem{lovelock}D.~Lovelock, J.\ Math.\ Phys.\  {\bf 12 }, 498-501 (1971).

\bibitem{nunes}
  S.~E.~Perez Bergliaffa and Y.~E.~C.~de Oliveira Nunes,
  Phys.\ Rev.\ D {\bf 84}, 084006 (2011)
  [arXiv:1107.5727 [gr-qc]].


\bibitem{CB}
T.~Clifton and J.~D.~Barrow,
Phys.\ Rev.\ D\ {\bf 72}, 103005  (2005)
[gr-qc/0509059].

\bibitem{Deser}
S.~Deser, O.~Sarioglu and B.~Tekin,
Gen.\ Rel.\ Grav.\ \ {\bf 40}, 1  (2008)
[arXiv:0705.1669 [gr-qc]].

\bibitem{Multamaki}
T.~Multamaki and I.~Vilja,
Phys.\ Rev.\ D\ {\bf 74}, 064022  (2006) [astro-ph/0606373];
T.~Multamaki and I.~Vilja,
Phys.\ Rev.\ D\ {\bf 76}, 064021  (2007) [astro-ph/0612775].

\bibitem{CapozzielloStabileTroisi}
S.~Capozziello, A.~Stabile and A.~Troisi,
Class.\ Quant.\ Grav.\  {\bf 25}, 085004 (2008)
[arXiv:0709.0891 [gr-qc]].

\bibitem{Bezerra}
T.~R.~P.~Carames, E.~R.~Bezerra de Mello,
Eur.\ Phys.\ J.\  {\bf C64}, 113-121 (2009)
[arXiv:0901.0814 [gr-qc]].

\bibitem{Saffari}
R.~Saffari and S.~Rahvar,
Phys.\ Rev.\ D\ {\bf 77}, 104028  (2008)
[arXiv:0708.1482 [astro-ph]];
Mod.\ Phys.\ Lett.\  A {\bf 24} 305 (2009)
[arXiv:0710.5635 [astro-ph]].

\bibitem{SSSsolutions} L.~Sebastiani and S.~Zerbini,
Eur.\ Phys.\ J.\  C {\bf 71}, 1591 (2011)
[arXiv:1012.5230 [gr-qc]].

\bibitem{Visser:1993nu}
M. Visser, Phys.\ Rev.\ D {\bf 48}, 5697 (1993).

\bibitem{D} S.~Deser and B.~Tekin,
Phys.\ Rev.\ D\ {\bf 67}, 084009  (2003) [hep-th/0212292];
S.~Deser and B.~Tekin, Phys.\ Rev.\ D  {\bf 75}, 084032 (2007) [arXiv:gr-qc/0701140].

\bibitem{Visser}G.~Abreu and M.~Visser,{\em Phys.\ Rev.\ Lett.}\  {\bf 105}, 041302 (2010).

\bibitem{Cai}R.~G.~Cai, L.~M.~Cao, Y.~P.~Hu and N.~Ohta, Phys.\ Rev.\ D  {\bf 80}, 104016 (2009) [arXiv:0910.2387 [hep-th]].

\bibitem{SSSEnergy}
G.~Cognola, O.~Gorbunova, L.~Sebastiani and S.~Zerbini,
Phys.\ Rev.\  D {\bf 84}, 023515 (2011)
[arXiv:1104.2814 [gr-qc]].

\bibitem{Wald} R.M. Wald,
Phys. Rev. D {\bf 48}, 3427 (1993).

\bibitem{HT} S. W. Hawking, Nature {\bf 248} 30 (1974);
Commun. Math. Phys. {\bf 43} 199-220 (1975).

\bibitem{Vanzo}L.~Vanzo, Phys.\ Rev.\ D {\bf 56}, 6475 (1997) [gr-qc/9705004].

\bibitem{r}R.~J.~Riegert, Phys.\ Rev.\ Lett.\  {\bf 53}, 315-318 (1984).

\bibitem{m}P.~D.~Mannheim and D.~Kazanas, Astrophys.\ J.\  {\bf 342}, 635-638 (1989).

\bibitem{klem}D.~Klemm, Class.\ Quant.\ Grav.\  {\bf 15}, 3195-3201 (1998) [gr-qc/9808051].

\bibitem{m2}
P.~D.~Mannheim, Astrophys.\ J.\  {\bf 479}, 659 (1997) [astro-ph/9605085];
Prog.\ Part.\ Nucl.\ Phys.\  {\bf 56}, 340-445 (2006) [astro-ph/0505266].

\bibitem{habib}
  S.~Habib Mazharimousavi, M.~Halilsoy and T.~Tahamtan,
  Eur.\ Phys.\ J.\ C {\bf 72}, 1851 (2012)
  [arXiv:1110.5085 [gr-qc]].


\bibitem{FaraoniEntropy}
V.~Faraoni,
Entropy {\bf 12}, 1246 (2010)
[arXiv:1005.2327 [gr-qc]].

\bibitem{VisserHaw}
M.~Visser,
Int.\ J.\ Mod.\ Phys.\ D\ {\bf 12}, 649  (2003)
[hep-th/0106111].


\bibitem{jac}
  T.~Jacobson and G.~Kang,
  Class.\ Quant.\ Grav.\  {\bf 10}, L201 (1993)
  [gr-qc/9307002].

\bibitem{nada}
  M.~Nadalini, L.~Vanzo and S.~Zerbini,
  Phys.\ Rev.\ D {\bf 77}, 024047 (2008)
  [arXiv:0710.2474 [hep-th]].

\bibitem{PW}
M.~K.~Parikh and F.~Wilczek, Phys.\ Rev.\ Lett.\  {\bf 85}, 5042 (2000).

\bibitem{Nadalini}
M.~Angheben, M.~Nadalini, L.~Vanzo and S.~Zerbini,
JHEP {\bf 0505}, 014 (2005)
[arXiv:hep-th/0503081].
M.~Nadalini, L.~Vanzo and S.~Zerbini,
J.\ Phys.\ A  {\bf 39}, 6601 (2006)
[arXiv:hep-th/0511250].

\bibitem{remo}
  D. Christodoulou and R.~Ruffini,
  Phys.\ Rev.\ D {\bf 4}, 3552 (1971).

\bibitem{Pope}
  H.~Lu, Y.~Pang, C.N.~Pope and J.F.~Vazquez-Poritz,
  Phys.\ Rev.\ D {\bf 86}, 044011 (2012)
  [arXiv:1204.1062 [hep-th]].

\bibitem{de}
  A.~De Felice and S.~Tsujikawa,
  Living Rev.\ Rel.\  {\bf 13}, 3 (2010)
  [arXiv:1002.4928 [gr-qc]].




\end{thebibliography}
\end{document}